\documentclass[preprint,5p,twocolumn,sort&compress]{elsarticle}
\usepackage{mathtext,bbm,amsmath,amsfonts,amssymb,indentfirst,syntonly,graphicx}
\usepackage[english]{babel}
\usepackage{amsmath,amssymb,amsthm}
\usepackage{latexsym,graphicx,bbm}

\newcommand{\beq}{\begin{eqnarray}}
\newcommand{\eeq}{\end{eqnarray}}

\newcommand{\ba}{\left( \begin{array}}
\newcommand{\ea}{\end{array} \right)}

\begin{document}
\title{A unified constraint on the Lorentz invariance violation from both short and long GRBs}
\date{\today}
\author[ihep,tpcsf]{Zhe Chang }
\ead{changz@ihep.ac.cn}

\author[ihep,tpcsf]{Yunguo Jiang \corref{cor1}}
\ead{jiangyg@ihep.ac.cn}

\author[ihep,tpcsf]{Hai-Nan Lin}
\ead{linhn@ihep.ac.cn}

\cortext[cor1]{Corresponding author at: Institute of High Energy Physics, Chinese Academy of Sciences, 100049 Beijing, China.}
\address[ihep]{Institute of High Energy Physics, Chinese Academy of Sciences, 100049 Beijing, China}
\address[tpcsf]{Theoretical Physics Center for Science Facilities, Chinese Academy of Sciences, 100049 Beijing, China}

\begin{abstract}
Possible  Lorentz invariance violation (LIV) has been investigated for a long time based on observations of GRBs. These arguments relied on the assumption that  photons with different energy are emitted at the same place and time. In this work, we try to take account of the intrinsic time delay $\Delta t_{\rm int}$ between emissions of low and high energy photons by using the magnetic jet model. The possible LIV effects are discussed in a unified scenario both for long and short {\it Fermi}-detected GRBs. This leads to a unique quantum gravity energy scale $M_1c^2 \sim 1.0 \times 10^{20}$ GeV  respecting the linear dispersion relation.
\end{abstract}

\maketitle

\section{Introduction}

Lorentz invariance is one of the most important cornerstones of modern physics. Recently, the OPERA collaboration reported that the GeV neutrinos  propagate faster than the speed of light \cite{Adam:2011zb}. Although other independent experiments should be performed to verify the superluminal phenomenon, it is valuable to ask whether Lorentz invariance violation (LIV) happens in high energy scale. A favored way to test LIV is to study the most explosive events in the present universe: gamma-ray bursts (GRBs). The {\it Fermi}~~satellite has observed several GRBs with photon energy $>100 $ MeV in recent years. {\it Fermi}~~carries two instruments: the Gamma-Ray Burst Monitor (GBM) and the Large Area Telescope (LAT), which detect the energy band $8 {\rm KeV}- 40 {\rm MeV} $ and $30 {\rm MeV}-300 {\rm GeV}$, respectively. An interesting feature of the observation is that  GeV photons arrive several seconds later than  MeV photons \cite{Abdo2009a,Abdo2009b,Abdo2009c,Ackermann2011}.

One possible explanation is given by quantum gravity effects. Some quantum gravity theories predict that high energy photons may interact with the
foamy structure of the space-time, thus photons with different energy propagate with different velocities \cite{Gambini:1998it,Ellis:2008gg,Ellis:2009vq,Ellis:2011ek}.
Such effects can be accumulated after photons travel a cosmological distance. In these theories, high energy photons are subluminal.
But there are still some other theories which show that high energy photons can be superluminal. For example, LIV can also be induced from the geometry of the space-time itself, such as the Finsler geometry \cite{Chang:2011td,Li:2011ia,Kostelecky:2011qz}. One can expect that the velocity of photon also depends on the energy, and may be superluminal.

A straight forward way to test LIV is studying the individual GRB, several papers have discussed the upper limits of variations of the light speed \cite{Schaefer:1998zg,Boggs:2003kxa}. Ellis et al. have proposed a data fitting procedure to test LIV effects \cite{Ellis:2011ek,Ellis:2005wr,Ellis:1999sd,Shao:2009bv}. The linear fitting function is expressed as $\Delta t_{\rm obs} /(1+z) =a_{\rm LIV} K(z)+b$, where $K(z)$ is a non-linear function of  the redshift measuring the cosmological distance, and $b$ represents the ignorance of the intrinsic time lag. A statistical ensemble of GRBs was used to fit values of $a_{\rm LIV}$ and $b$, and no strong evidence of LIV was found. The explicit form of $K(z)$ depends on the cosmological model. Biesiada and Pi\'{o}rkowska applied this procedure to various cosmological models \cite{Biesiada:2009zz}. Shao et al. used this method to discuss four {\it Fermi}-detected GRBs  \cite{Shao:2009bv}, which we will discuss in the present work. Nevertheless, all these investigations concentrated on the time lag induced by LIV, the intrinsic time lag which depends on the emission mechanism of GRBs was not considered.

On the other hand,  without considering LIV effects, several  mainstream GRB models  were proposed to explain the delayed arrival of GeV photons \cite{Beloborodov:2009be,Meszaros2011,Bosnjak:2011pt,Duran2011}. M\'esz\'aros and Rees presented a magnetic dominated jet model to explain this phenomenon. The MeV photons can escape the plasma when their optical depth decreases to unity at the photosphere radius. While the GeV photons are produced by the nuclear collision between  protons and neutrons, which happens at a large radius (compared to the photosphere radius) \cite{Meszaros2011}. Duran and Kumar considered that photons are emitted by  electrons via the synchrotron radiation, it consumes more time for  electrons to be accelerated in order to radiate GeV photons \cite{Duran2011}. Bo$\check{\rm s}$njak and Kumar proposed the magnetic jet model, the time delay depends linearly on the distance where the jet is launched \cite{Bosnjak:2011pt}.

In this work, we argue that  the observed time lag for two photons with energy $E_{\rm high}$ and $E_{\rm low}$ consists of two parts,
\beq \Delta t_{\rm obs} = \Delta t_{\rm LIV} + \Delta t_{\rm int},  \label{eq:flyint} \eeq
where $\Delta t_{\rm int}$ denotes the intrinsic emission time delay, and $\Delta t_{\rm LIV}$ represents the flying time difference  caused by LIV effects.  In section \ref{sec:mdj}, we  take use of  the magnetic jet model in Ref.\cite{Bosnjak:2011pt} to estimate $\Delta t_{\rm int}$.  The LIV induced time lag $\Delta t_{\rm LIV}$ is given, and the quantum gravity energy scale is discussed in Sec. \ref{sec:liv}. Finally, the discussion and conclusions are given  in Sec. \ref{dis}.

\section{Magnetic jet model \label{sec:mdj}}

In the magnetic jet model, photons with energy less than $10$ MeV can escape when the jet radius is beyond the
Thomson photosphere radius, i.e., the optical depth for low energy  photons is $\tau_{T} \sim 1$. However,  GeV photons will be converted to electron-positron pairs at this radius, and can escape later when the pair-production optical depth $\tau_{\gamma \gamma}(E)$ drops below unity.

The bulk Lorentz factor of an expanding spherical fireball increases with the radius roughly as $\Gamma\propto r$, until reaching a saturate radius $r_s$ where the Lorentz factor is saturated \cite{Meszaros2006}. However, for an expanding jet with small ejecting angle, the effective dynamical dimension is one. The bulk Lorentz factor increases with the radius roughly as \cite{Bosnjak:2011pt}
\begin{eqnarray}
  \Gamma(r)\approx\begin{cases}
   ( r/r_0)^{1/3}\quad \ & \ {\rm for}\quad r_0 \lesssim r\lesssim r_s,\\
    \eta\quad \ & \ {\rm for}\quad r\gtrsim r_s,\\
  \end{cases}
\end{eqnarray}
where  $r_0\approx 10^7$ cm is the base of the outflow, which represents the distance from the central engine where the jet is launched. $\eta$ is the final bulk Lorentz factor of the jet.

The optical depth for the photon-electron scattering is defined as
\begin{equation} \label{optdept}
  \tau_T(r)=\int_r^{\infty} \frac{dr'}{2\Gamma^2}\sigma_T n \Gamma,
\end{equation}
where $\sigma_T=e^4/( 6\pi\varepsilon_0^2c^4m_e^2) \approx 6.65\times10^{-25}$ cm$^2$ is the Thomson scattering cross-section. The baryon number density in the observer frame is $n \simeq L/4\pi r^2m_p\Gamma c^3\sigma_0$, where $L$ is the isotropic luminosity, $m_p$ is the mass of protons, $\sigma_0 \equiv \Gamma(r_0)[1+\sigma(r_0)]$, and $\sigma(r_0)$ is the initial ratio of the magnetic  and  baryon energy densities.
When $\tau_T(r)=1$, the Thomson photon sphere radius is
\begin{equation} \label{tomoptdep}
\frac{r_p}{r_0}\approx 1.36 \times 10^5 L_{52}^{3/5} \sigma_{0,3}^{-3/5} r_{0,7}^{-3/5},
\end{equation}
where $L_{52}\equiv L/10^{52}\,\, {\rm erg}\cdot {\rm s}^{-1}$, $\sigma_{0,3} \equiv \sigma_0/10^3$, and  $r_{0,7}\equiv r_0/10^7$ cm. We use  cgs units for numerical values here and after.

The optical depth for a photon of energy $E_0$ to be converted to $e^{\pm}$ while traveling through the jet at a radius $r$, is given by \cite{Bosnjak:2011pt}
\begin{equation}
  \tau_{\pm}(E_0,r)=\bigg{(}\frac{\beta-2}{\beta-1}\bigg{)}\frac{\sigma_{\gamma\gamma}}{4\pi r \Gamma^2}\frac{L_{>p}}{(1+z)^{3-2\beta}E_p c}\bigg{[}\frac{E_pE_0}{\Gamma^2m_e^2c^4}\bigg{]}^{\beta-1},
\end{equation}
where  $\sigma_{\gamma\gamma}=6\times 10^{-26}$ cm$^2$ is the cross-section for photons producing $e^{\pm}$ just above the threshold energy, $E_p$ is the peak energy of the $\nu F_{\nu}$ spectrum, $L_{>p}$ is the frequency-integrated luminosity above $E_p$, and $\beta\approx2.2$ is the photon index of the spectrum  above  $E_p$.
Setting $\tau_{\pm}=1$, the pair-production photonsphere radius is given by
\begin{equation} \label{eq:rgg}
  \frac{r_{\gamma\gamma}(E_0)}{r_0}\approx 4.13 \times10^6L_{>p,52}^{0.41}E_{p,-6}^{0.08}E_{0,-4}^{0.49}r_{0,7}^{-0.41}(1+z)^{0.57} ,
\end{equation}
 where $E_{p,-6}=E_p/{\rm MeV}$, and $E_{0,-4}=E_0/100{\rm MeV}$.

In the observer frame, the relative time delay  between MeV photons and GeV photons equates to the time for the jet to propagate from $r_p$ to $r_{\gamma\gamma}$ \cite{Bosnjak:2011pt}
\begin{equation}
  \Delta t=\frac{3r_0(1+z)}{2c}\bigg{[}\bigg{(}
  \frac{r_{\gamma\gamma}(E_0)}{r_0}\bigg{)}^{1/3}-\bigg{(}\frac{r_p}{r_0}\bigg{)}^{1/3}\bigg{]}.
\label{dt} \end{equation}
Taking use of Eq.(\ref{dt}) together with Eq.(\ref{tomoptdep}) and  Eq.(\ref{eq:rgg}), one can calculate the time delay for the arrival of GeV and 100 MeV photons relative to MeV photons. Thus, the intrinsic time delay is $\Delta t_{\rm int}=\Delta t(E_{\rm high})-\Delta t(E_{\rm low})$. The observed values of $E_p$, $E_{\rm ios,54}$, T$_{90}$ and $z$ for four $Fermi$-detected GRBs are taken from \cite{Bosnjak:2011pt}, and are listed in Table \ref{observedvalue}.

\begin{table}
\begin{center}
\begin{tabular}[t]{ccccc}
\hline\hline
GRB & E$_p$ & E$_{\rm ios,54}$ &T$_{90}$ &z\\
 &keV &erg &s & \\
\hline
080916c &424 & 8.8 & 66 &4.35\\
090510 &3900 &0.11 & 0.6 &0.90\\
090902b &726 &3.7  &22  &1.82\\
090926 &259 &2.2 &13 &2.11\\
\hline
\end{tabular}
\end{center}
\caption{\small{The observed parameters of four $Fermi$-detected GRBs. $E_p$ is the energy at the peak of $\nu f_{\nu}$ spectrum. $E_{\rm ios,54}$ is the isotropic equivalent energy in unit of $10^{54}$ ergs. T$_{90}$ is the GRB duration which $90$\% of the counts are above background. $z$ is the GRB redshift \cite{Bosnjak:2011pt}. }}
\label{observedvalue}
\end{table}

\section{Test of LIV effects \label{sec:liv}}

As is mentioned above, some quantum gravity theories predict that  photons with different wavelength propagate in different speed \cite{Gambini:1998it, Ellis:2008gg, Ellis:2009vq, Schaefer:1998zg,Biesiada:2009zz,AmelinoCamelia:1997gz}. The non-trivial space-time structure may affect the propagation of photons, so high energy photons may arrive later than low energy ones.  Many works have studied  LIV effects of the high energy photons in this direction \cite{Ellis:2011ek,Ellis:2005wr,Nemiroff:2011fk,Ellis:2002in,Jacob:2008bw,Shao:2010wk}.

Consider two photons emitted at the same time and place, within the LIV phenomenology, the arrival time delay between them can be written as \cite{Nemiroff:2011fk,Jacob:2008bw}
\beq \Delta t_{\rm LIV} = \frac{1+n}{2 c} \big(\frac{\Delta E}{M_n c^2}\big)^n  D_n, \eeq
where $n=1$ or $n=2$ denotes the linear or quadratic correction to the dispersion relation, and $D_n$ is defined to be \cite{Nemiroff:2011fk,Jacob:2008bw}
\beq D_n \equiv \frac{c}{H_0} \int_0^{z} \frac{(1+z')^n dz'}{\sqrt{\Omega_{M}(1+z')^3+\Omega_{\Lambda}}}, \label{eq:cosdistance} \eeq
where $H_0 \simeq 72 $ km sec$^{-1}$ Mpc$^{-1}$ is the Hubble constant, $\Omega_{M}$ and $\Omega_{\Lambda}$ are the present values of the matter density and cosmological constant density, respectively. In the standard cosmological model, ($\Omega_{M}, \Omega_{\Lambda}$) are given by observations as ($0.3,0.7$) \cite{Jacob:2008bw}. For $n=1$, the time delay depends linearly on the variation of energy, which we will consider in the follow. In this case, the effective LIV energy scale is
\beq  M_1c^2 = \frac{\Delta E D_1}{ c \Delta t_{\rm LIV}}. \eeq

In Ref.\cite{Shao:2009bv}, Shao et al. took use of 4 GRBs the same as in Table \ref{observedvalue} to predict LIV effects, the observed time delay $\Delta t_{\rm obs}/(1+z)$ vs. $K(z)$ was plotted, where $K(z)$ is defined as
 \beq K(z) \equiv \frac{\Delta E}{(1+z)} \frac{D_1}{c}. \eeq
In their plot,  the three long bursts  were found to be near one line.  However, the short burst was not fitted well by the same line. The intercept of the line was interpreted as $\Delta t_{\rm int}/(1+z)$, and the slope of the line can be interpreted as $1/M_1c^2$.   $\Delta t_{\rm int}/(1+z)$ was found to be negative, which means that high energy photons are emitted earlier than  low energy ones. This conflicts with  standard GRB models.
Taking into account of the four GRBs, the quantum gravity energy scale is estimated to be $M_1c^2\sim 2\times 10^{17}$ GeV for the linear energy dependence. In their work, the intrinsic time delay is assumed to be the same for three long bursts. Besides, their result strongly depends on the artificial choices. For example, if the 33.4 GeV photon is replaced by the 11.16 GeV photon in GRB 090902b, the three long bursts can not be fitted well by one line.

In the follow, we first calculate  $\Delta t_{\rm int }$ by using Eq.(\ref{dt}). Then, combining with the observation data, we give $\Delta t_{\rm LIV}$ by the relation given in Eq.(\ref{eq:flyint}). In Ref.\cite{Nemiroff:2011fk}, the authors used a statistical method to determine the
observed time difference $\Delta t_{\rm obs}$ between photons with energy $E_{\rm low}$ and $E_{\rm high}$. In our present work, $E_{\rm low}$ is taken to be 100 MeV, since the onset of $100$ MeV photons can be read directly from the data of the  LAT monitor. $E_{\rm high}$ is the energy of  the most energetic photon in each GRB. One exception is that the second energetic photon with $E_{\rm high}=11.16$ GeV in GRB 090902b is chosen. The most energetic $33.4$ GeV photon arriving at $82$ s is excluded due to its incoincidence with the main burst. In Ref.\cite{Shao:2009bv}, Shao et al. selected this event to estimate LIV effects without considering the central engines and emission mechanism. In the magnetic jet model, if we believe that the delayed GeV photons are due to the high optical depth, the arrival time of the $33.4$ GeV photon should be 3 s later than the 11.16 GeV photon. However, the observed time interval $70$ s  is far beyond the model's prediction. This photon may be due to the  inelastic collision of protons and neutrons \cite{Beloborodov:2009be,Meszaros2011}, and it is quite possible that this individual event happens when the jet encounter the interstellar medium.

 With Eq.(\ref{dt}), we can estimate  $\Delta t_{\rm int } \simeq 0.06$ s for GRB 090510, if $r_0=10^6$ cm. Then $\Delta t_{\rm LIV} \simeq 0.14$ s,
 and $M_1c^2\sim 9.73 \times 10^{19}$ GeV, which is about 8 times of the Planck energy $E_{\rm Planck} \sim 10^{19}$ GeV \cite{Abdo2009c}.  However, if we increase $r_0$  to $10^7$ cm, then $\Delta t_{\rm int}\simeq 0.46$ s and $\Delta t_{\rm LIV}\simeq -0.26$ s. In this case, high energy photons become superluminal, which is against the argument of quantum gravity theories. It is a reasonable assumption that $r_0\simeq 10^6$ cm for GRB 090510, because this is a short burst and its radius should be small than long bursts.

\begin{table*}
\begin{center}
\begin{tabular}[t]{ccccccc}
\hline\hline
GRB & $E_{\rm low}$ &$E_{\rm high}$&$\Delta t_{\rm obs}$   &$\Delta t_{\rm LIV}$ & $K(z)$ & $M_1c^2$ \\
&MeV &GeV &s &s & s$\cdot$GeV &GeV  \\
\hline
080916c  &100 & 13.22 & 12.94 & 0.24 &4.50 $\times10^{18}$& 10.02 $\times10^{19}$  \\
090510 &100&31 &0.20 &0.14    &7.02 $\times10^{18}$&  9.73 $\times10^{19}$  \\
090902b &100 &11.16 & 9.5  &0.10  &3.38 $\times10^{18}$& 9.94 $\times10^{19}$ \\
090926 &100 &19.6 &21.5 &0.20&6.20 $\times10^{18}$&9.59 $\times10^{19}$ \\
\hline
\end{tabular}
\end{center}
\caption{\small{The LIV induced time delay $\Delta t_{\rm LIV}$ and quantum gravity energy scale $M_1c^2$ derived from four {\it Fermi}-detected GRBs. $\Delta t_{\rm obs}$ is collected from Ref.\cite{Abdo2009a,Abdo2009b,Abdo2009c,Ackermann2011}. $\Delta t_{\rm LIV}=\Delta t_{\rm obs}-\Delta t_{\rm int}$, where $\Delta t_{\rm int}$ is calculated by Eq.(\ref{tomoptdep}), Eq.(\ref{eq:rgg}) and Eq.(\ref{dt}). The value of $\sigma_{0,3}$ in each GRB is approximately the bulk Lorentz factor of the jet in unit of $10^3$ and is taken as $\sigma_{0,3}\sim 1$ \cite{Bosnjak:2011pt}. $r_{0,7}$ is chosen as 16.7, 0.1, 28.7 and 55.0 for GRB 080916c, GRB 090510, GRB 090902b and GRB 090926, respectively.} }
\label{timedelay} \end{table*}

\begin{figure}
\centering
  \includegraphics[width=8cm]{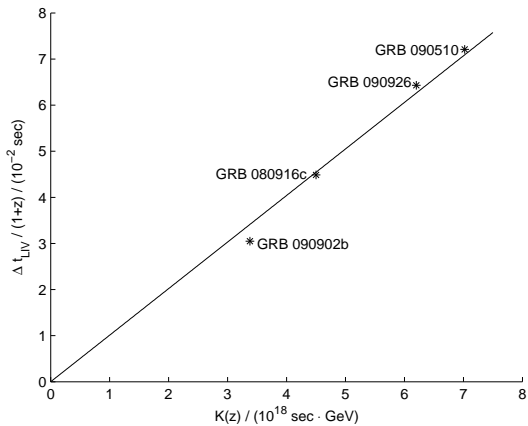}\\
  \caption{The  plot of $\Delta t_{\rm LIV}/(1+z)$ vs. $K(z)$ for four {\it Fermi}-detected GRBs.}\label{fig}
\end{figure}

With the above criteria, we try to use the line fitting method to predict the LIV effects. In principle, if the linear dispersion relation holds, the $\Delta t_{\rm LIV}/(1+z)$ vs. $K(z)$ plot should be a zero-intercept line, whose slope is the inverse of quantum gravity energy scale, i.e., $1/M_1c^2$. By choosing $r_0$ of each burst properly, the limits of LIV effects for both short and long bursts can be unified respecting the linear dispersion relation.  The four GRB points can be fitted well by one line, if we choose $r_{0,7}=$ 16.7, 0.1, 28.7 and 55.0 for GRB 080916c, GRB 090510, GRB 090902b and GRB 090926, respectively. In this case, LIV effects are calculated in Table \ref{timedelay}, and the  $\Delta t_{\rm LIV}/(1+z)$ vs. $K(z)$  plot is given in Fig.\ref{fig}. The values of $r_0$, which indicate the active scale of central engines, are reasonable.  The inverse of the slope gives $M_1c^2\sim 1.0\times 10^{20}$ GeV, which is roughly the same result of the GRB 090510. The energy scale of the modified photon dispersion relation is one order of magnitude higher than the conventional Planck scale. This may suggest that the quantum gravity scale may be more subtle than one naively thinks, because this quantity is model dependent. For instance, the effective quantum energy scale depends on the density of $D$-particles in the D-foam model \cite{Ellis:2008gg,Ellis:2009vq}. It is an interesting future work to combine both the quantum gravity model and the GRB model together to study the LIV effects. Supposing LIV effects are strongly suppressed, which is the assumption taken by the mainstream GRB models, the observed time delay can well predict the value of $r_0$ by using the magnetic jet model, since the tuning of $r_0$ is sensitive to the fitting.

\section{Discussion and conclusion \label{dis}}

From Eq.(\ref{dt}), one can infer that the time lag has approximate linear relation with the initial radius ($\propto r_0^{0.86}$) and the redshift ($\propto (1+z)^{1.19}$). It has a weak dependence on the photon energy $E_0$, the peak energy $E_p$, and the peak luminosity $L_{>p}$, which are $\Delta t_{\rm int} \propto E_0^{0.17}, E_p^{0.05},$ and $L_{>p}^{0.14}$, respectively. In addition, the time lag is hardly dependent on $\sigma_0$.
Another model proposed to explain the delayed GeV photons was given by M\'esz\'aros and Rees. In this model, the time lag also depends linearly
on $r_0$ (See Eq.(19) in \cite{Meszaros2011}).
These GRB models also predict the spectrum, so they can be verified by the observed spectrum of the whole energy band.

As mentioned above, in Ellis et al.'s proposal,  the linear fitting function is written as $\Delta t_{\rm obs} /(1+z) =a_{\rm LIV} K(z)+b$. In our case, $b \sim \Delta t_{\rm int}/(1+z)$ is not a constant, and we can estimate it approximately as
\beq b \simeq 0.08\, r_{0.7}^{0.86}L_{>p,52}^{0.14} E_{p,-6}^{0.03}E_{0,-4}^{0.16}(1+z)^{0.19}, \label{eq:b} \eeq
which roughly agrees with  Ellis et al.'s result $b \sim 10^{-2}$ \cite{Ellis:2005wr}. Since $b$ weakly depends on the redshift,  it can be regarded as a distance-independent quantity \cite{Shao:2009bv}. However, $b$ strongly depends on $r_0$, so the long and the short bursts will lead to quite different values of $b$. If the long bursts have roughly the same $r_0$, $b$ can be considered as a constant, and the linear relation between $\Delta t_{\rm obs} /(1+z) $ and $K(z)$ holds.

The analysis above gives us the hint that, if we want to consider source effects of the GRBs, long bursts with small redshifts are preferred. Due to the short cosmological distance, quantum gravity effects can be attenuated.   With the purpose of enhancing quantum gravity effects, short GRB bursts with high redshifts should be selected. Therefore, the future observations on short GRBs will improve the test of LIV effects.
If we use an ensemble of GRBs with both long and short bursts, the fitting function method is not convincing. The intrinsic time delay plays an important role. Better knowledge of the intrinsic property of the source will help us to improve the test of LIV effects. Inversely, better understanding of quantum gravity can help us to predict the parameters in GRB models.

In this work, we discussed LIV effects by making use of the magnetic jet model.  GeV  photons are emitted later than MeV photons, due to their different optical depths. This physical ingredient should be included in probe of LIV effects. The neglect of the photon emission mechanism may lead to misleading results. The constraints of LIV effects can be unified for both long and short bursts.  The calculation of the linear energy dependence of dispersion relation gives $M_1c^2 \sim 1.0 \times 10^{20}$ GeV. Although the magnetic jet model  itself should be tested by further investigations, the analysis of the intrinsic time delay is important when we study the photons from the astrophysical sources to test LIV effects.

\section*{Acknowledgments}
We are grateful to M. H. Li, X. Li and S. Wang for useful discussion. This work has been funded in part by the National Natural Science Fund of China under Grant No. 10875129 and No. 11075166.


\end{document}